# Combining randomized field experiments with observational satellite data to assess the benefits of crop rotations on yields

Dan M. Kluger[a], Art B. Owen[a], David B. Lobell[b]

[a] Department of Statistics, Stanford University

[b] Department of Earth System Science and Center on Food Security and the Environment, Stanford University

**Abstract:**

With climate change threatening agricultural productivity and global food demand increasing, it is important to better understand which farm management practices will maximize crop yields in various climatic conditions. To assess the effectiveness of agricultural practices, researchers often turn to randomized field experiments, which are reliable for identifying causal effects but are often limited in scope and therefore lack external validity. Recently, researchers have also leveraged large observational datasets from satellites and other sources, which can lead to conclusions biased by confounding variables or systematic measurement errors. Because experimental and observational datasets have complementary strengths, in this paper we propose a method that uses a combination of experimental and observational data in the same analysis. As a case study, we focus on the causal effect of crop rotation on corn (maize) and soy yields in the Midwestern United States. We find that, in terms of root mean squared error, our hybrid method performs 13% better than using experimental data alone and 26% better than using the observational data alone in the task of predicting the effect of rotation on corn yield at held-out experimental sites. Further, the causal estimates based on our method suggest that benefits of crop rotations on corn yield are lower in years and locations with high temperatures whereas the benefits of crop rotations on soy yield are higher in years and locations with high temperatures. In particular, we estimated that the benefit of rotation on corn yields (and soy yields) was 0.84 t/ha (0.23 t/ha) on average for the top quintile of temperatures, 1.02 t/ha (0.20 t/ha) on average for the whole dataset, and 1.18 t/ha (0.15 t/ha) on average for the bottom quintile of temperatures.

## Section 1: Introduction

The task of understanding which farm management practices lead to increased crop yields has been important for millennia. This task is especially critical in a time with increasing rates of food insecurity [1] and with climate change threating agricultural productivity [2]. Increasing crop yields with better farm management practices can mitigate food-insecurity issues and can substantially decrease the amount of land used for agriculture—a sector which currently uses about 37% of the land on Earth [3]. Yet defining which practices are truly "better" for a given location and climatic regime can be very difficult given the typically strong interactions of management with soil and weather conditions.

Historically, researchers have turned to randomized experiments on designated research croplands to answer questions about whether a particular agricultural practice leads to higher crop yields. Such randomized experiments are the gold standard in causal inference because they can be used to get an unbiased estimate of the causal effect of a farm management practice on yield. On the other hand, randomized field trials often suffer from small sample sizes leading to wider confidence intervals than desired for these causal effects. In addition, of even greater concern is that randomized experiments can suffer from *external validity* issues [4], meaning that their conclusions may not apply to farms with different soil, climate, or management conditions than those of the experimental site. Recent open science initiatives that aggregate the data from many different field experiments [5, 6, 7], can reduce small sample size and external validity issues; however, even in aggregated datasets with a variety of field experiments there may be few experiments where the management practice of interest is randomized, and these experiments could still suffer from external validity issues.

Thanks to the recent big data revolution coupled with advances in satellite-based remote sensing technologies, the effects of agricultural practices on yields are now being investigated with much larger datasets that span a wide array of growing conditions [8, 9, 10]. Despite the appeals of using large satellite-based datasets for drawing causal inferences, they suffer from two main drawbacks. First, satellites can give inaccurate estimates of treatment variables such as crop rotation or tillage as well as inaccurate estimates of the yield, although this issue will be of decreasing importance as the technology and algorithms continue to improve. Second, such datasets are observational and lack randomized treatments, so causal inferences drawn from such datasets would rely heavily on the assumption that there are no unmeasured confounders. If strong assumptions fail to hold, an observational study could fail to satisfy *internal validity*, meaning that the estimates for the causal effect based on the observational study are biased and do not reflect the true causal effect in the study sample.

The advantages and drawbacks of experimental and observational datasets are complimentary, suggesting benefits for combining the two types of datasets in the same analysis. In particular, it is plausible to leverage high internal validity of experimental data to improve upon causal estimates from observational studies which could have high biases due to systematic measurement errors or unmeasured confounders. Conversely, it is plausible to leverage the large sample size and high external validity of observational data to improve upon causal estimates from experimental data, which could have high variance due to small sample sizes and may not be representative of the fields and growing conditions of interest. Statistical methods for fusing both experimental and observational data is a growing area of research [11], and a number of recent theoretical results from the field of statistics suggest that when both experimental and observational data are available, it is strictly better to use certain methods that leverage both the experimental and observational data in tandem rather than to only use the experimental data [12, 13] or to only use the observational data [14]. To our knowledge, these hybrid methods have not previously been used in agronomy or the earth sciences.

The goal in this paper is to develop an approach for combining data from randomized field experiments with observational satellite-based datasets to estimate the causal effect of agricultural practices on yields. As a case study, our focus is on estimating the causal effect of crop rotation on corn and soy yields in the Midwestern United States.

Crop rotation has long been known to be an important component of agricultural systems and its benefits on crop yields are well established via randomized experiments. For example, for the predominant two-year corn soy rotation in the United States, the benefit of rotation on corn yields was 0.80 tons per hectare (t/ha) on average and ranged from -0.32 to 1.68 t/ha (5%-95% quantiles) across 28 field experiments conducted prior to 2008 [15]. The benefit of rotation on soy yields is similarly well established [16, 17, 18, 19].

The yield benefit of rotation is commonly explained by reductions in pest or disease pressure, as well as by increased nutrient availability and soil quality. Less is understood about how the effectiveness of crop rotation varies across geographical regions and weather conditions and about how the effectiveness of crop rotation might change under the extreme temperature and precipitation conditions that are forecasted in a changing climate. To this end, interactions between crop rotation benefits and weather or soil covariates have been explored using solely experimental data [20] or solely observational data [10, 9]. The results in [20] suffer from small sample sizes and suggest that the rotation benefit on corn yields is higher at low temperatures whereas the results from [10, 9] do not use data from randomized experiments and suggest the opposite relationship between rotation benefit on corn yields and temperatures.

In this paper, we introduce a calibration-based approach for combining both experimental and observational data in the same analysis. We validate that the approach is helpful for predicting the treatment effects at held-out experimental sites. Finally, we explore interactions between our estimated crop rotation benefits and weather covariates and discuss our findings.

## Section 2: Methods

### Section 2.1: Study area and timespan

The Corn Belt of the United States is characterized by high-yielding commercial agriculture for corn and soybeans and contributes nearly 30% of the global production for these crops [22]. We focus on a 9-state region within the Corn Belt (Figure 1) spanning the 19-year period from 2000-2018 due to availability of the yield data (Section 2.2.1).

### Section 2.2: Datasets

#### Section 2.2.1: Satellite-based observational dataset

Our observational dataset consisted of 20,000 randomly sampled 30m x 30m agricultural pixels from each of the 9 states in our study region, resulting in a dataset of 180,000 pixels. Each pixel's distance to its nearest neighbor in our sample was generally much larger than 100m (median=875m, 5%-95% quantile= [205m, 2315m]), implying our samples were reasonably well spread out and did not contain many repeat observations from the same plot of land.

The yield data in our observational dataset was taken from previously published yield maps for corn [23] and soybean [24] that were produced using the Scalable Crop Yield Mapper (SCYM). SCYM, initially developed in [25] and further improved in [26, 23, 24], is an approach that uses both the crop yield and the leaf area index time series output by regionally parameterized crop growth simulations to fit a statistical model that predicts crop yield with both remotely sensed vegetation indices and weather variables as covariates. In [23], the produced SCYM-based corn yield map was compared to a large ground dataset and was found to have county level agreement of $r^2$=0.67 (RMSE= 1.10 t/ha) and pixel (30m x 30m) level agreement of $r^2$=0.40 (RMSE= 2.45 t/ha). In [24], the produced SCYM-based soybean yield map had county level agreement of $r^2$=0.63 (RMSE= 0.40 t/ha) and pixel level agreement of $r^2$=0.27 (RMSE= 0.96 t/ha) with the ground data.

For each 30m x 30m pixel in our dataset, crop type was determined using the Corn-Soy Data Layer [27] for years prior to 2008 and the Cropland Data Layer [28] for years since 2008, inclusive. The 2-year window crop rotation was inferred from this data and was categorized into five types of observed rotations: corn planted before corn (CC), soy planted before corn (SC), soy planted before soy (SS), corn planted before soy (CS), and rotations involving other crops. Our analysis ignores rotations involving other crops, because the specific crop type among the other crops could not be determined with accuracy that was close to that of the corn and soy classifications and because yield maps were not available for the other crops. Table S1 gives the number of pixels observed for each of the five rotation categories as well as the average SCYM yield for that rotation category.

Using Google Earth Engine, daily temperatures were extracted from PRISM [29, 30] and precipitation variables were extracted from a ~4 km resolution meteorological dataset [31]. Soil properties for the top one meter were extracted from the SSURGO soil database [32]. To simplify our analysis, we pre-selected a small number of weather and soil covariates that are known to be associated with crop yields and, in combination, were thought to be highly correlated with the dropped weather and soil variables. A list of our pre-selected weather and soil covariates along with their definitions and sources can be found in Table 1. The procedure used to calculate growing degree days (GDD) and extreme degree days (EDD) is described in Appendix A. GDD and EDD are temperature metrics constructed based on findings in [33] such that high GDD is thought to be beneficial to corn growth while high EDD is thought to be detrimental to corn growth.

**Table 1:** Pre-selected weather and soil covariates and their descriptions. Each row corresponds to an element of the feature vector *X* described in Section 2.3.1.

| Covariate name | Description | Measurement frequency | Source |
|---|---|---|---|
| Latitude | Latitude of 30m x 30m pixel | once | |
| Longitude | Longitude of 30m x 30m pixel | once | |
| Year | Year in which measurements were taken | annually | |
| Growing degree days (GDD) | Aggregated temperature exceeding 8° C (April-September) | annually | PRISM [29, 30] |
| Extreme degree days (EDD) | Aggregated temperature exceeding 30° C (April -September) | annually | PRISM [29, 30] |
| Early season precipitation | Rainfall between January 1st and April 30th | annually | GRIDMET [31] |
| Growing season precipitation | Rainfall between May 1st and September 15th | annually | GRIDMET [31] |
| Previous year early season precipitation | Early season precipitation from the previous year | annually | GRIDMET [31] |
| Previous year growing season precipitation | Growing season precipitation from the previous year | annually | GRIDMET [31] |
| rootznaws | Rootzone available water storage | once | SSURGO [32] |
| aws0_100 | Top 1 meter available water storage | once | SSURGO [32] |
| Corn productivity index (NCCPI [34]) | Score of how favorable the soil is for growing corn | once | SSURGO [32] |
| Soy productivity index (NCCPI [34]) | Score of how favorable the soil is for growing soy | once | SSURGO [32] |

### Section 2.2.2: Experimental dataset

Our experimental dataset came from two open-source datasets [5, 6], each containing the yield data from multiple different randomized field experiments in North America. Some sites were included in both datasets, although in those cases we only considered the data from [6] because it included more years of data. For experiments measuring corn yield, we removed all experiments that either (i) did not use any fertilizer, (ii) were located more than 50 km away from our 9-state study region, or (iii) did not contain instances of both CS and CC. For experiments measuring soy yield, we had the same inclusion criteria except instead of (iii), we required that the experiment had instances of both CS and SS. We further removed experimental data from before the year 2000 and subplots that did not use fertilizer from our analysis. Our final dataset consisted of 11 experiments, each lasting at least three years, and combined they included 89 site-years of yield data and spanned the years 2000-2016. All experiments in our final dataset were completely randomized block designs with 2 to 5 replications per site [35, 36, 37, 38, 21].

In Figure 1, we see that the 11 experiments in our final dataset were spread out throughout the Corn Belt and were conducted in a variety of climatic conditions. Despite this variety, our dataset contained no experiments from certain states, only one experiment in a climate with low early season precipitation, and only one experiment in a climate with a large number of extreme degree days.

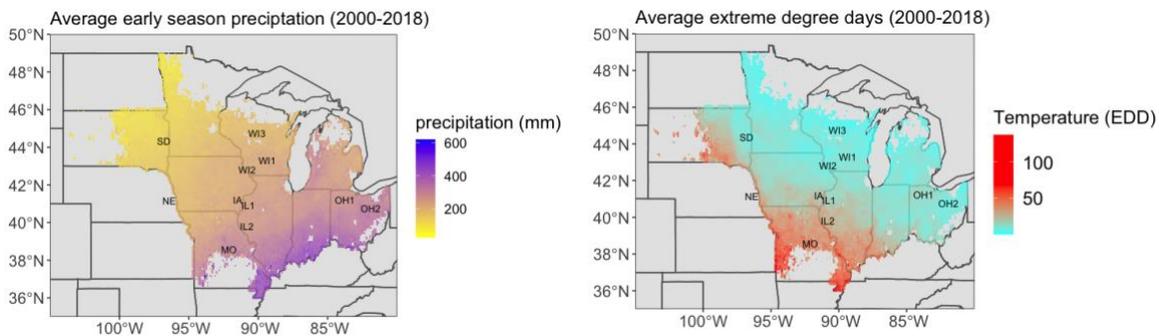

**Figure 1:** In these maps the colored regions indicate the geographical span of the satellite-based dataset (Section 2.2.1), and the text indicates the location of the 11 experiments (Section 2.2.2). The map on the left shows the variation in geography of early season precipitation, averaged across the years 2000-2018 while the map on the right shows the geographical variation in extreme heat, measured in terms of extreme degree days averaged across the years 2000-2018.

**Section 2.3: Estimation of the treatment effects**

**Section 2.3.1: Estimation of the treatment effect using satellites only**

To estimate the heterogenous treatment effect of crop rotation on yield using our satellite-based observational data only, we fit a causal forest [39, 40], using the `grf` R package [41]. The causal forest method is a recent adaptation of the classical random forest algorithm [42] designed for estimation and inference of causal effects when the treatment effects are heterogenous. It has previously been used to estimate the effect of tillage on crop yields using a combination of satellite, weather, and soil data [8]. More broadly, the causal forest method has been used to evaluate the impact of weather on agricultural productivity [43] as well as effectiveness of forest management policies [44], fishery policies [45], and growth mindset interventions [46].

To explain the inputs of our causal forest, the quantities we estimated with it, and important assumptions for using a causal forest, we introduce some notation and definitions. Let Yield_{Corn,2} be the yield of corn in the second year of a two-year window. Let Rot denote the rotation in a two-year window. Finally, let $X$ denote a vector of our chosen covariates, which contains year, geolocation, weather covariates and soil covariates (see Table 1). Letting $x$ be a specific realization of these covariates, we apply a causal forest to our satellite-based dataset to nonparametrically estimate the following function of $x$:

$$\omega_{\text{Corn}}(x) = \mathbb{E}(\text{Yield}_{\text{Corn},2}|X=x, \text{Rot}=\text{SC}) - \mathbb{E}(\text{Yield}_{\text{Corn},2}|X=x, \text{Rot}=\text{CC}).$$

In words, $\omega_{\text{Corn}}(x)$ is the mean second-year corn yield in a soy-corn rotation minus that in a corn-corn rotation, stratified by the specific year, geolocation, soil covariates and weather covariates encoded in $x$. Throughout this text, we will refer to the estimate, output by the causal forest, of the above function as ECRB_{Sat}(x), which stands for the satellite-based estimated corn rotation benefit. We remark that difficulty estimating $\omega_{\text{Corn}}(x)$ due to failure to meet the overlap assumption is not a concern in our setting (see Figure S1).

It should be emphasized that ECRB_{Sat}(x) is not a reliable estimate of the true causal effect of rotation on corn yields for two reasons. First, $\omega_{\text{Corn}}$ is only guaranteed to be the true causal effect of rotation under the assumption that there are no unmeasured confounders. This assumption is unlikely to hold because fertilizer is well known to increase crop yields and because farmers that rotate their crops likely use less fertilizer in order to reduce operational costs (in fact, the economically optimal fertilization rate is consistently much lower for SC than for CC [47], and such estimates are well-advertised to farmers). Since fertilizer is neither measured nor included in our covariate vector $X$, we expect $\omega_{\text{Corn}}(x)$ to be an underestimate of the true effect of crop rotation on corn yields. Second, ECRB_{Sat}(x) is an estimate of $\omega_{\text{Corn}}(x)$ using a satellite-based dataset which has errors in both the measured rotation type and the yield. We expect the errors in the measured rotation will lead the causal forest to underestimate the absolute value of $\omega_{\text{Corn}}(x)$ (see Appendix B for a mathematical justification), and the errors in the SCYM-based yield may further bias the estimate of $\omega_{\text{Corn}}(x)$. Taken together, these two issues imply that ECRB$_{\text{Sat}}(x)$ is likely to be a biased estimate of the true causal effect of rotation on corn yield.

Analogously, we define Yield$_{\text{Soy,2}}$ to be the yield of soy in the second year of a two-year window and

$$\omega_{\text{Soy}}(x) = \mathbb{E}(\text{Yield}_{\text{Soy,2}}|X = x, \text{Rot} = \text{CS}) - \mathbb{E}(\text{Yield}_{\text{Soy,2}}|X = x, \text{Rot} = \text{SS}).$$

We also fit a causal forest on the CS and SS in our satellite-based dataset to estimate $\omega_{\text{Soy}}(x)$. We call this estimate ESRB$_{\text{Sat}}(x)$, which stands for the satellite-based estimated soy rotation benefit. For similar reasons, we expect ESRB$_{\text{Sat}}(x)$ to be a biased estimate for the true causal effect of rotation on soy yield.

We caution that based on Figure S1, estimates of $\omega_{\text{Soy}}(x)$ corresponding to observations with high propensity scores are likely to be somewhat inaccurate. Often this issue is addressed by removing observations with propensity scores near 0 and 1 [8, 48], restraining the causal analysis to a particular subpopulation where both treatments and controls are commonly observed. In the main text of this paper, we do not remove observations based on propensity scores; however, in Appendix C we show that our results are not sensitive to this decision to keep points with high propensity scores that could be prone to inaccurate estimates of $\omega_{\text{Soy}}(x)$.

**Section 2.3.2: Estimation of the treatment effect using experiments only**

We estimated the experimental treatment effect of rotation on corn yield by pairing the yield data. In particular, we paired SC subplots with CC subplots, such that in each pair, the two subplots were within the same site, year and replicate, and also had the same tillage, fertilizer, and drainage level. Because our focus was on rotation effects under the most common management practices, we note that we dropped paired observations from subplots with zero fertilizer (this only happened in the Nebraska site) as well as subplots that were in more complex rotations that happened to contain a two-year SC sequence (e.g. the Corn-Oats-Wheat-Soy rotation in the site from South Dakota was dropped). In each site $s$ and year $t$, we estimated the effect of rotation on corn yield for the $j$th pair by taking the difference between the yields in the two subplots:

$$\text{ECRB}_{\text{Exp}}(s, t, j) = \text{Yield in SC subplot - yield in CC subplot.}$$

For each site $s$, $\text{ECRB}_{\text{Exp}}(s, t, j)$ was calculated for all years between 2000-2018, which had available yield data and which were at least one year after the rotations at that site were initiated.

Similarly, for each site $s$ and year $t$, we estimated the experimental treatment effect of rotation on soy yields for each subplot pair $j$ with

$$\text{ESRB}_{\text{Exp}}(s, t, j) = \text{Yield in CS subplot - yield in SS subplot.}$$

Because most sites did not have yield data for a continuous soy rotation, we were only able to calculate $\text{ESRB}_{\text{Exp}}(s, t, j)$ at three out of the 11 experimental sites in our study.

**Section 2.3.3: Hybrid approach for estimating the treatment effect**

To produce a heterogenous estimate of the causal effect of rotation on yield, which leverages both the experimental datasets and the satellite-based observational dataset, we fit a linear calibration of the $\text{ECRB}_{\text{Sat}}$ values at the experimental sites towards the $\text{ECRB}_{\text{Exp}}$ values. In particular, at each experimental site $s$ and year $t$, we used the precise latitudes and longitudes of the experimental plots at each site to extract the weather and soil covariates at that site from the same data sources in Section 2.3.1 [29, 30, 31, 32]. Using the covariate vector $x(s,t)$ from each site $s$ and year $t$, we computed $\text{ECRB}_{\text{Sat}}(x(s,t))$ using the fitted causal forest (Section 2.3.1). Using all the site and year combinations that had both $\text{ECRB}_{\text{Sat}}(x(s,t))$ and $\text{ECRB}_{\text{Exp}}(s, t, j)$ estimates, we fit the following mixed effects regression model

$$\text{ECRB}_{\text{Exp}}(s, t, j) = a + b \times \text{ECRB}_{\text{Sat}}\big(x(s, t)\big) + \alpha_s + \beta_t + \varepsilon_{s,t,j},$$

where $\alpha_s$ is a random effect for site, $\beta_t$ is a random effect for year, and $\varepsilon_{s,t,j}$ is mean zero, independent Gaussian noise. The intercept and slope terms were included in the model because, as suggested in Section 2.3.1, we expect ECRB$_{\text{Sat}}$ values to have both additive and multiplicative biases as estimates of the true causal effect, and we wish to estimate the additive and multiplicative biases in order to correct for them. ECRB$_{\text{Exp}}$ values are thought to be unbiased estimates of the treatment effect but they have quite high variance. The random effects for the site were included to account for varying management practices at the different experimental sites. The random effects for the year were included to account for the possibility that the bias in ECRB$_{\text{Sat}}$ could change from year to year (this can happen if the unmeasured confounders such as fertilizer rate vary from year to year).

After fitting the mixed effects model, we used the estimated intercept $\hat{a}$ and slope $\hat{b}$ to construct a calibrated estimate of the effect of crop rotation on corn yield for any choice of covariate $x$:

$$\text{ECRB}_{\text{Calib}}(x) = \hat{a} + \hat{b} \times \text{ECRB}_{\text{Sat}}(x).$$

To calibrate the benefit of rotation on soy yields, we only have 3 experimental sites with continuous soy yield data (each with only 4 or 5 years of data), so the above approach of using a mixed effects model to estimate the calibrated rotation benefit would be unreliable on the soy yield data. Instead, we simply try to correct for the additive bias in $\text{ESRB}_{\text{Sat}}(x)$, because we do not have enough distinct experimental sites to correct for the multiplicative bias as well. In particular, we set

$$\text{ESRB}_{\text{Calib}}(x) = \overline{\text{ESRB}_{\text{Exp}}(s,t,j)} - \overline{\text{ESRB}_{\text{Sat}}(x(s,t,j))} + \text{ESRB}_{\text{Sat}}(x),$$

where $\overline{\text{ESRB}_{\text{Exp}}(s,t,j)}$ is the average experimental effect observed across all pairs of rotated soy and continuous soy while the site-wise average experimental effects and $\overline{\text{ESRB}_{\text{Sat}}(x(s,t,j))}$ is the average satellite-based estimates of the soy rotation benefit across all pairs of rotated soy and continuous soy.

## Section 2.4: Leave one out validation of rotation benefit estimates

To estimate the errors associated with our calibration approach in predicting the experimental rotation effect on corn yield at an unobserved experimental site, we used a variant of leave one out cross-validation. In particular, for each of the 11 experimental sites, we first dropped that site, then estimated the calibration parameters $\hat{a}$ and $\hat{b}$ using the 10 remaining sites based on the regression in Section 2.3.3. We then used these fitted parameters coupled with the $\text{ECRB}_{\text{Sat}}(x(s,t))$ values at the held-out experimental site to predict the average experimental effect $\text{ECRB}_{\text{Exp}}$ across the years in which the held-out experiment occurred. Finally, we computed the squared error between the predicted average experimental effect in the held-out site and the observed average value of $\text{ECRB}_{\text{Exp}}$ at the experimental site. We repeated this process for all 11 experimental sites and computed the square root of the mean of the aforementioned squared prediction errors.

To demonstrate the benefits of using both experimental and observational data, rather than merely experimental data we repeated the leave one out validation for various modifications of the mixed effects model in Section 2.3.3. In particular, we considered dropping the observational $\text{ECRB}_{\text{Sat}}(x(s,t))$ terms entirely or replacing $\text{ECRB}_{\text{Sat}}(x(s,t))$ with weather covariates. We also considered predicting held-out sites using merely the effect observed in the nearest experimental site geographically rather than the model in Section 2.3.3. This allowed us to estimate the benefit of using the observational data rather than solely using the experimental data in predicting the experimental effect at an unobserved experimental site. As a sensitivity check, we repeated this analysis when including a tillage versus no tillage indicator as a predictor variable and when removing the random site and time effects $\alpha_s$ and $\beta_t$ to set up standard linear regression.

The leave one out validation was only performed for predicting the experimental rotation benefit on corn and not for predicting the experimental rotation benefit on soy because the latter only had 3 experimental sites.

## Section 3: Results and Discussion

### Section 3.1: Satellite-based estimates, experiment-based estimates and their association

The satellite-based and experimental estimates for the benefit of crop rotation on both corn yield and soy yield, calculated using the methods of Sections 2.3.1 and 2.3.2, are visualized in Figures S2 and S3. As expected, we generally see a positive effect of rotation on both corn and soy yield, although the observational corn rotation benefit, $ECRB_{Sat}$, is negative in some regions. While there could be cases where the true causal effect of rotation on corn yield is negative, we suspect that in many of the regions where $ECRB_{Sat}$ is negative, the uncalibrated estimates are simply giving the wrong sign.

To directly visualize how correlated the satellite-based estimates are with the experiment-based estimates, we create a scatterplot of $ECRB_{Sat}(x(s,t))$ versus the average of $ECRB_{Exp}(s,t,j)$ across the $j$ subplot pairs (Figure 2, top) as well as a scatterplot of $ESRB_{Sat}(x(s,t))$ versus the average of $ESRB_{Exp}(s,t,j)$ across the $j$ subplot pairs (Figure 2, bottom). Each point in the scatter plot corresponds to one year of data at one site.

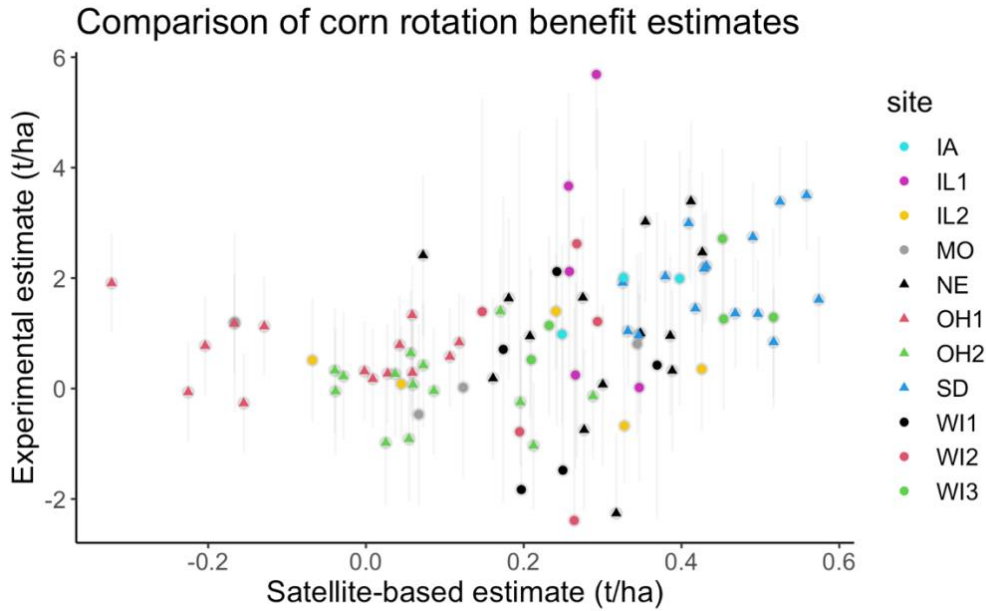

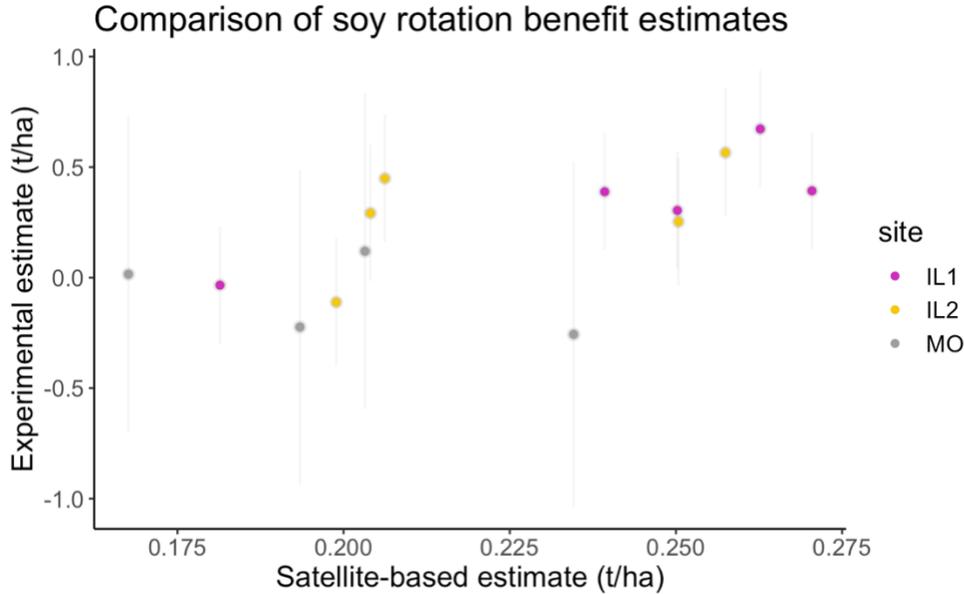

**Figure 2:** Top: Each point corresponds to one year and one experimental site. The *x* coordinate gives the ECRB$_{Sat}$ value. The *y* coordinate gives the value of ECRB$_{Exp}$ averaged across all paired subplots within a particular experimental site and year, with 95 percent confidence intervals for the year and site-specific averages given by the grey lines. The 95 percent confidence interval pooled standard error estimates across years but not across sites. Triangular points represent sites from [6] with 14-15 years of data while circular points represent sites from [5] with 3-5 years of data. Bottom: This is similar to the top panel except here we look at estimates for the effect of rotation on soy yields rather than those for corn yields.

We fit the model described in Section 2.3.3 and estimated $\hat{a} = 0.58$ (95% CI = [0.00,1.15]) and $\hat{b} = 1.75$ (95% CI= [0.37,3.12]). These confidence intervals are based on the estimated standard errors in the mixed effects model, and the positive confidence interval for $b$ gives evidence that there is a positive association between ECRB$_{Sat}$ and ECRB$_{Exp}$. To double check the statistical significance of the positive association between ECRB$_{Sat}$ and ECRB$_{Exp}$ we ran 10,000 iterations of the cluster bootstrap, where the experiments were sampled with replacement, but the data within each experiment was not (see for example, "Strategy 1" in Section 3.8 of [49]) . The 95 percent bootstrap confidence interval for the Pearson correlation between ECRB$_{Sat}$ and ECRB$_{Exp}$ was [0.03, 0.36] and the 95 percent bootstrap confidence interval for the linear regression slope of ECRB$_{Exp}$ on ECRB$_{Sat}$ was [0.33, 4.11]. The positive association observed between ECRB$_{Sat}$ and ECRB$_{Exp}$ is a reassuring sign that the satellite-based approach (Section 2.3.1) provides reasonable estimates of the true causal effect of rotation on yields.

For soy calibration, the additive bias correction term $\overline{\text{ESRB}_{\text{Exp}}(s,t,j)} - \overline{\text{ESRB}_{\text{Sat}}(x(s,t,j))}$ was equal to 0.003 t/ha (95% CI= [-0.10,0.11]). Therefore, our calibrated estimates $\text{ESRB}_{\text{Calib}}$ and $\text{ESRB}_{\text{Sat}}$ are nearly identical, and some readers may prefer to interpret our plots of $\text{ESRB}_{\text{Calib}}$ as simply plots of $\text{ESRB}_{\text{Sat}}$ values.

## Section 3.2: Leave one out validation

The results for the leave one out validation described in Section 2.4 indicate that the hybrid calibration approach is better at predicting the treatment effect in an unobserved experimental site than various reasonable approaches to using only the experimental data (Table 2). We also found that the hybrid calibration is better at predicting the treatment in an unobserved experimental site than using only the observational data. Overall, the hybrid approach resulted in errors that were 13% lower than using experimental data alone, and 26% lower than using the satellite data alone. The second and third rows of Table 2 indicate that these findings are not sensitive to our calibration model choices of including a random effect for site and year and our choice of not including tillage in the model. In Table S2, we consider an additional sensitivity check to see if our conclusions would still hold had we taken a weighted average rather than a standard average of the squared prediction errors when preforming leave one out cross validation.

**Table 2:** Root mean squared error (t/ha) when predicting the average experimental effects at held-out sites. In columns 4-9, values were calculated by fitting the model in Section 2.3.3 when the $\text{ECRB}_{\text{Sat}}(x(s,t))$ term is dropped and is either not replaced (column 4) or replaced by one weather covariate (columns 5-8) or replaced by all four weather covariates in a multivariate regression (column 9). In the first row, the preferred mixed effects model is used. The second row includes a sensitivity checks where we drop the random effects for site and year from the calibration model and fit a linear model instead. The third row includes another sensitivity check where we add an indicator of whether each experimental subplot had tillage as a variable in the random effects calibration model. The entries with the lowest root mean square error in each row are in bold.

|  | Satellite only | Experiment only | | Experiment plus weather covariates | | | | | Hybrid |
| --- | --- | --- | --- | --- | --- | --- | --- | --- | --- |
|  | Just predict with $\text{ECRB}_{\text{Sat}}$ | Nearest experiment | All other experiments | Early season precipitation | growing season precipitation | GDD | EDD | All four weather covariates | Calibration approach (Section 2.3.3) |
| Preferred model | 0.99 | 0.97 | 0.84 | 0.77 | 0.83 | 0.88 | 0.83 | 0.83 | **0.73** |
| Linear model | 0.99 | 0.97 | 0.84 | 0.75 | 0.84 | 0.92 | 0.87 | 0.82 | **0.73** |
| Tillage model | 0.99 | 1.24 | 0.92 | 0.88 | 0.92 | 0.96 | 0.92 | 0.93 | **0.84** |

These tables suggest that the hybrid calibration approach of Section 2.3.3 performs the best overall in predicting the treatment effect at a held-out site, but it does not perform substantially better than fitting a model that uses just the experimental data and the early season precipitation weather covariate. That being said, the hybrid calibration approach was a principled choice in the absence of a priori knowledge about which weather covariates are associated with the treatment effect. In contrast, the model with early season precipitation was merely found to perform well empirically, among four possible choices of weather covariates, on a dataset with a small number of different experiments.

### Section 3.3: Positivity and heterogeneity of estimated treatment effects

Having established that the calibration leads to improved estimates of the effect of rotation on corn yields, we visualize a map of these calibrated effects in Figure 3 (top). We also visualize $ESRB_{Calib}$ in Figure 3 (bottom), but we emphasize that because we have only 3 experiments with continuous soy data, our calibration merely accounted for additive bias rather than both additive bias and multiplicative bias, and the approach was not validated on held-out data. Figure 3 indicates that treatment effects of rotation on yield were found to be geographically heterogenous. The benefit of crop rotation on corn yield appears larger in the northwest parts of the Corn Belt, whereas the benefit of crop rotation on soy yield appears largest in the southern and central parts of the Corn Belt.

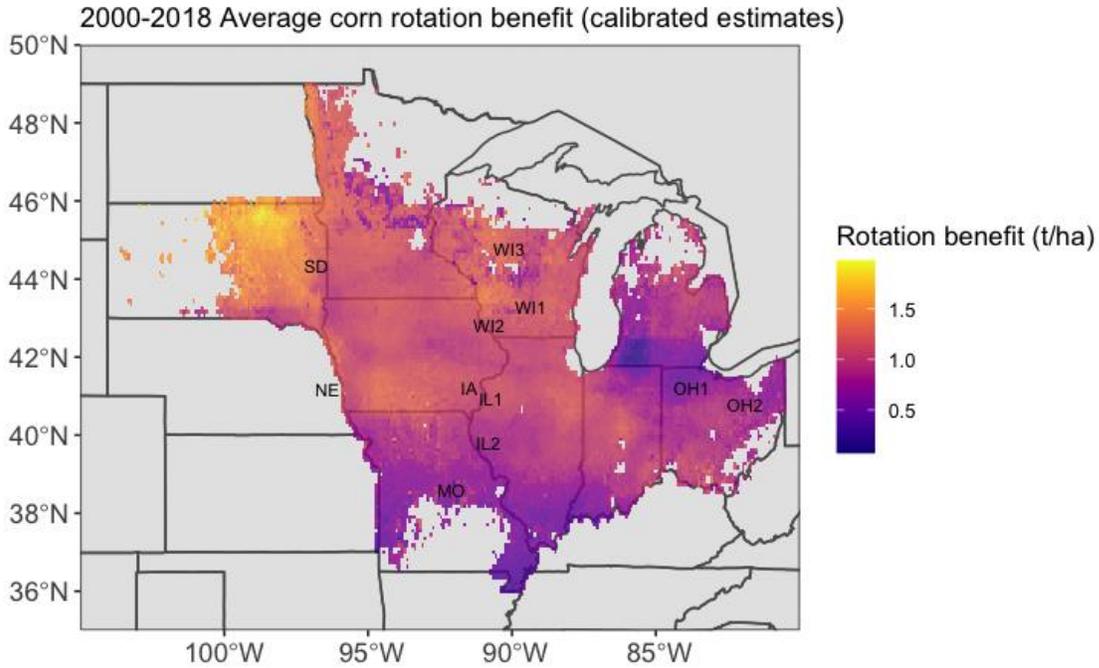

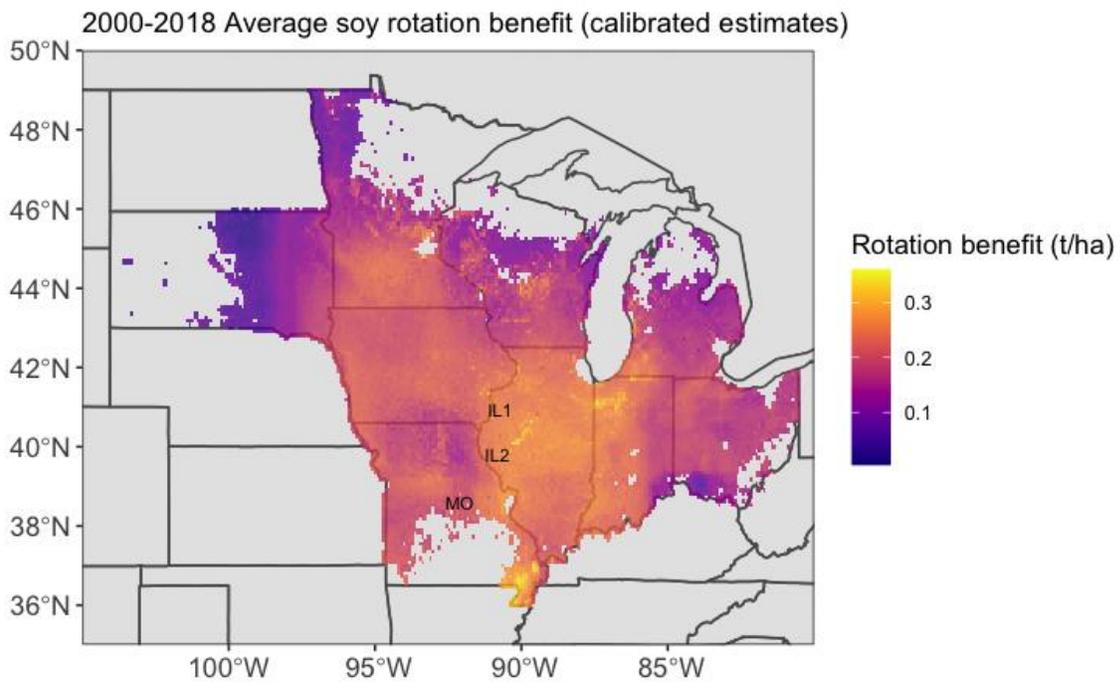

**Figure 3:** Top: ECRB$_{Calib}$ averaged across the years 2000-2018 and across 10 km$^2$ square bins. Bottom: ESRB$_{Calib}$ averaged across the years 2000-2018 and across 10 km$^2$ square bins. The text on these maps indicates the locations of experimental sites.

We can also see from Figure 3 that the calibrated rotation benefits for corn yield are generally positive. While only 89.1% of $ESRB_{Sat}$ values are positive, 99.6% of $ESRB_{Calib}$ values are positive (one sided 95 percent CI=[88%,100%], using the same cluster bootstrap as in Section 3.1), implying that the causal effect of rotation on corn yields is negative less often than observational data alone would suggest.

A key benefit of using observational data in addition to experimental data is the ability to study the treatment effect under weather conditions that were not observed in the experimental data. In Figure 4, we plot a heatmap of our calibrated estimates of the rotation benefit for soy and corn yields as a function of temperature and precipitation, with points indicating the weather covariates observed at experimental sites.

We observe that the corn rotation benefit is smaller at high temperatures and low growing season precipitation values, but this likely could not be inferred from experimental data alone as only one site-year was observed to have temperature greater than 2,400 growing degree days and growing season precipitation less than 500 mm. The lower absolute effects of rotation on corn yield in the high temperature and low precipitation regime are perhaps partly explained by these conditions leading to lower corn yields in general [50, 51, 52]. Heatmaps of $ECRB_{Calib}$ as a function of other temperature and precipitation variables are presented in Figure S4. While it appears that the benefit of rotation on yield is particularly high when at the lowest observed early season precipitation values, this phenomenon only occurs in a narrow window of temperature values, and is potentially an artifact of most of such points being in South Dakota, where the rotation benefit is seen to be particularly high. This phenomenon may also be explained by the greater uncertainty and inaccuracy of our estimates near the boundary of covariate space, which we discuss below.

We caution readers that near the boundaries of the heatmaps (e.g. outside 1$^{st}$-99$^{th}$ percentile range for the covariates) in Figures 4, S4 and S5, the estimated rotation benefits have higher uncertainty and are likely less accurate than those well within the interior of the heatmaps for two reasons. First, at each point near the boundary of covariate space, there are fewer samples within a neighborhood of that point, so the causal forest model will return $ECRB_{Sat}$ or $ESRB_{Sat}$ values that have higher variance. Second, the points near the boundary of covariate space are further away from the experimental sites, so if the true relationship between the satellite-based rotation benefits and the experimental-based rotation benefits is not the linear model described in Section 2.3.3, the extrapolation errors of using this calibration model will be largest near the boundary.

For soy we observe a different and clearer pattern (Figure 4, bottom). When the number of growing degree days is high, the rotation benefit for soy yields is high and the opposite is true when the number of growing degree days is low. Intriguingly, this pattern is not observed when we measure temperature with extreme degree days rather than growing degree days (Figure S5), and in fact for a fixed value of growing degree days, the benefit of rotation on soy yields does not increase with extreme degree days.

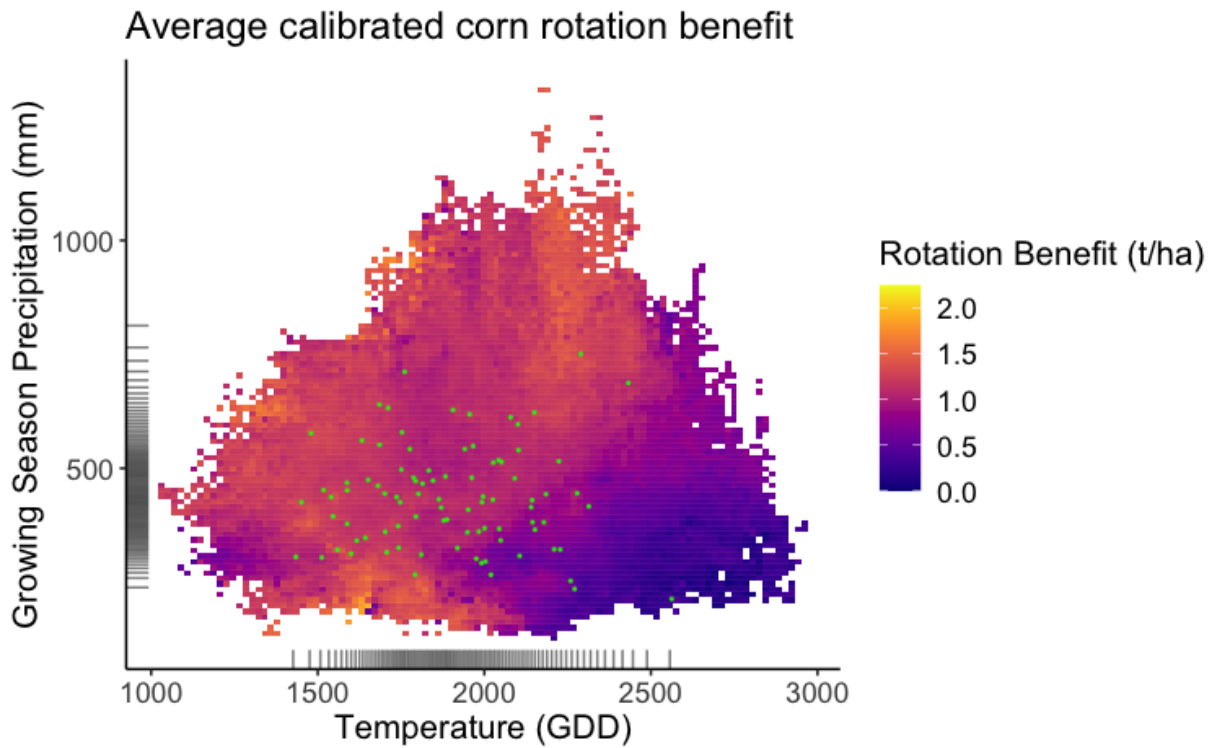

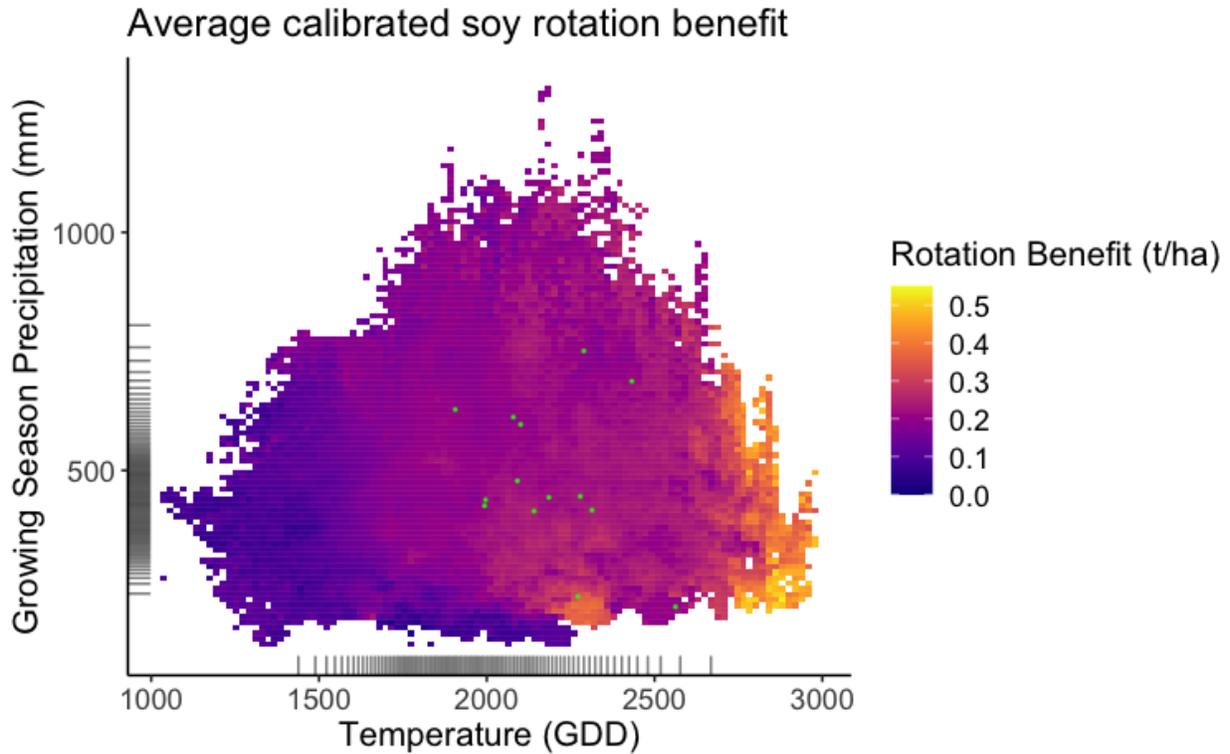

**Figure 4:** Top: Heatmap indicating how the calibrated corn rotation benefit, $ECRB_{Calib}$, varies with temperature and precipitation. Bottom: Heatmap indicating how the calibrated soy rotation benefit, $ESRB_{Calib}$, varies with temperature and precipitation. In both heatmaps, the green dots indicate the weather covariates of the experimental sites, with one dot per site per year. The 99 tick marks on each axes denote the $1^{st}$-$99^{th}$ percentiles of the corresponding weather covariate. As a complimentary analysis to the heatmaps in Figure 4, we compute the average calibrated rotation benefits in the top and bottom quintile for temperature (GDD). The average calibrated corn rotation benefit was 0.84 t/ha versus 1.18 t/ha across observations in the top and bottom quintile of GDD values, respectively. For soy yields, we observe the opposite pattern where rotation benefit is larger with larger temperatures. In particular, the average calibrated soy rotation benefit was 0.23 t/ha versus 0.15 t/ha across observations in the top and bottom quintile of GDD values, respectively. For reference, across the whole dataset, the average calibrated corn rotation benefit was 1.02 t/ha and the average calibrated soy rotation benefit was 0.20 t/ha.

We also check whether there is clear temporal trend in rotation benefits during the 19-year study period. In Figure S6, we see that there is no clear or substantial temporal trend in the rotation benefits for either corn or soy. It appears that $ECRB_{Calib}$ has a slight positive trend while the $ESRB_{Calib}$ values may have a slightly negative trend throughout the duration of the study period, a phenomenon which would be explainable by decreasing temperatures in the Corn Belt between 2000-2019.

**Section 3.4: Discussion**

Our findings on the corn rotation benefit are similar to those in [21]. In particular, they found that the benefit of crop rotation on corn yield persists in both drought and favorable conditions. We similarly find ECRB$_{Calib}$ to be positive across a range of conditions. Also, in most of their experimental sites they found a positive (but typically not statistically significant) interaction between rotation benefit and a particular corn growth favorability index (defined as the mean detrended corn yields), suggesting that rotations are less beneficial to corn yield in unfavorable growing conditions. Our analysis provides further evidence of this phenomena as we observe ECRB$_{Calib}$ to be lower in settings with high temperatures and low precipitation. Both analyses are in terms of absolute effects rather than percent effect, potentially explaining the lesser benefit of rotation on corn yield in unfavorable conditions. Finally, both analyses suggest a slight positive temporal trend (albeit not always a statistically significant one) in the benefit of rotation on corn yields in the Midwestern United States. It is also worth noting key differences between the analyses: ours incorporates weather covariates whereas theirs uses a mean detrended yield variable to determine favorability of weather conditions. In addition, our analysis only considers simple two-year corn soy rotations, whereas their analysis incorporates more complex rotations using a rotational diversity score. Future work can analyze complex crop rotations using more sophisticated diversity scores such as the one proposed in [53].

Pest pressure could largely explain our findings that the benefit of crop rotation on soy yields is larger in observations with higher temperatures (in GDD, but not in EDD). It is well established that the benefits of crop rotation for soy yields result largely because of reductions in pest pressure. For example, [54] found that the benefit of rotation on soy yields in two Louisiana experiments was largely explained by a reduction in soybean cyst nematode and [55] found crop rotation to be effective in reducing soybean cyst nematode. Meanwhile, using pesticide application rate as a proxy for pest pressure, [56] showed that increased yearly minimum temperatures are associated with increased pest pressure, as many agricultural pests are not able to survive low winter temperatures. Therefore, in warmer conditions, we would expect there to be higher pest pressure on soy, rendering crop rotation a more beneficial management practice in increasing soy yields because of its effectiveness in reducing pest pressure.

To explore the hypothesis that ESRB is larger in warmer conditions due to increased pest pressure, we plot pesticide use per acre versus ESRB values in Figure 5. Because pesticide use on soybean crops is thought to be a proxy for pest pressure on soybeans (pesticides tend to be applied after pests are reported in the region), the figure indicates that soy rotation benefit tends to be higher under conditions with greater pest pressure.

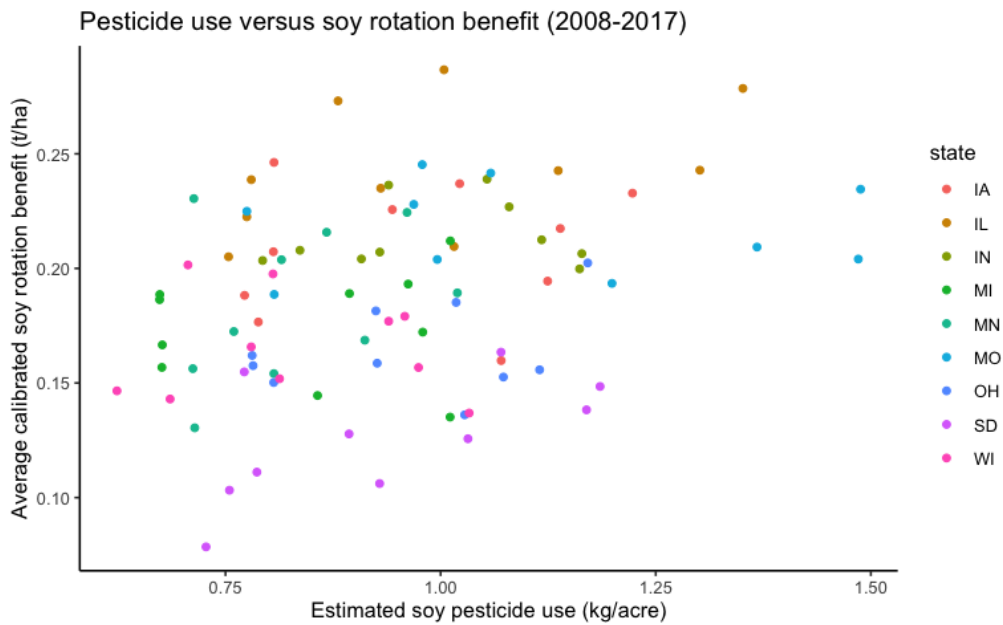

**Figure 5:** Plot of estimated pesticide use for soy versus average ESRB$_{Calib}$ values. Each point corresponds to one state and one year. To generate this plot, soybean acreage estimates were taken from NASS quickstats [57], and the high estimates of pesticide use were taken from the USGS [58]. We did not include pesticide use data from 2018 because this data was preliminary. We also did not include pesticide use data from prior to 2008 in Figure 5 because pesticide use did not appear to be associated with increased temperatures prior to 2008 (Figure S7) and because no direct measurements of pest pressure could be leveraged instead.

## Section 4: Conclusion

In this paper we developed an approach to combine experimental and observational datasets to assess the effect of agricultural practices on yields. The approach can be used to simultaneously mitigate external validity issues of randomized field experiments and internal validity issues of observational studies. Open-source initiatives which collated the results of many field experiments were essential to our approach. Further, more sophisticated approaches than our linear calibration approach could be worth considering in settings with many experimental sites.

Focusing on the causal effect of crop rotation on yields in the Midwestern United States as a case study, our hybrid approach led to better predictions of the causal effect at held-out experimental sites than did the standard approaches of using only experimental data or using only observational data (Tables 2 and S2). Furthermore, the case study highlighted some of the issues with observational data and experimental data. We found that using the observational data only led to a biased underestimate of the benefit of crop rotation on corn yields. This bias is partially explained by fertilizer being an unmeasured confounder because rotating fields tend to use less fertilizer. We also see that the experimental data alone do not adequately capture the geographical and weather scenarios of interest (Figures 1, 3, 4, S4 and S5).

Our results suggest that the benefit of rotation on corn yield is smaller in warmer conditions whereas the benefit of rotation on soy yield is larger in warmer conditions. The latter observation could be partly explained by increased pest pressure in warmer conditions. As we anticipate a warming climate, these findings have implications for how strongly crop rotations should be encouraged in the future. Specifically, our findings suggest that rotation may become a less beneficial management practice for corn yields in a future with higher temperatures, though rotation will still remain a net beneficial practice. Meanwhile, our findings suggest that rotation will become an increasingly important management practice for maximizing soy yields, and therefore, soy farmers should be increasingly encouraged to rotate their crops.

**Acknowledgements**

DMK was supported by the James and Nancy Kelso Stanford Interdisciplinary Graduate Fellowship and partially by a Stanford Graduate Fellowship. ABO was supported by NSF grant IIS-1837931. DBL was supported by the NASA Harvest Consortium (NASA Applied 787 Sciences Grant No. 80NSSC17K0652, sub-award 54308-Z6059203).

The authors gratefully acknowledge Jill Deines, Dominik Rothenhäusler, Michael Sklar, Matthieu Stigler, Paul Switzer, and Sherrie Wang for helpful discussions. The authors would also like to thank Lori Abendroth and Timothy Bowles, for providing relevant information about the experimental datasets that were used. Finally, the authors thank Jill Deines for assistance with extracting the observational data from Google Earth Engine.

# Supplementary Materials

**Table S1:** Our observational dataset had 180,000 pixels with yield and crop data at each pixel spanning 20 years each (1999-2018). The below table splits all 19 x 180,000 two-year windows for each pixel into 5 different two-year rotation categories and displays the number of each two-year windows observed for each rotation. The third column displays the average SCYM-based yield estimate observed in the second year of each two-year window.

| 2-year window crop rotation | Number of observations in pixel sample (1999-2018) | Average yield estimate of second crop in rotation |
|---|---|---|
| Corn planted before corn (CC) | 448,318 | 10.78 t/ha |
| Soy planted before corn (SC) | 864,150 | 11.01 t/ha |
| Soy planted before soy (SS) | 265,556 | 3.17 t/ha |
| Corn planted before soy (CS) | 881,078 | 3.49 t/ha |
| Other | 960,898 | NA |

**Table S2:** Root mean squared error (t/ha) when predicting the average experimental effects at held-out sites for the preferred mixed effects model. In the first row, the squared prediction error is averaged over each of the four long term sites (14-15 years), in the second row squared prediction error is averaged over the seven short term sites (3-5 years). In columns 4-9, values were calculated by fitting the mixed effects model in Section 2.3.3 when the $\text{ECRB}_{\text{Sat}}(x(s,t))$ term is dropped and is either not replaced (column 4) or replaced by one weather covariate (columns 5-8) or replaced by all four weather covariates in a multivariate regression (column 9). The entries with the lowest root mean square error in each row are in bold. We can see that had we assigned more weight to long term sites rather than simply taking an unweighted average of the squared prediction errors across all 11 experiments (as is done in the main text), the conclusion that the hybrid approach preforms better than approaches that use only experimental or observational data (but not both) still holds.

|  | Satellite only | Experiment only | | Experiment plus weather covariates | | | | | Hybrid |
|---|---|---|---|---|---|---|---|---|---|
|  | Just predict with $\text{ECRB}_{\text{Sat}}$ | Nearest experiment | All other experiments | Early season precipitation | growing season precipitation | GDD | EDD | All four weather covariates | Calibration approach (Section 2.3.3) |
| Long term sites | 0.93 | 0.79 | 0.78 | 0.60 | 0.82 | 0.70 | 0.81 | 0.64 | **0.58** |
| Short term sites | 1.01 | 1.06 | 0.88 | 0.85 | 0.84 | 0.97 | 0.84 | 0.92 | **0.81** |

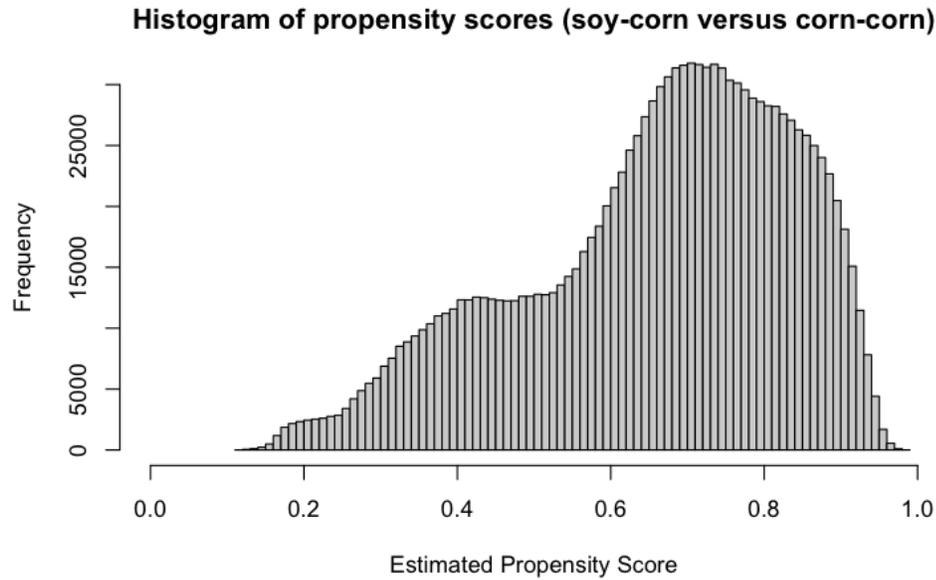

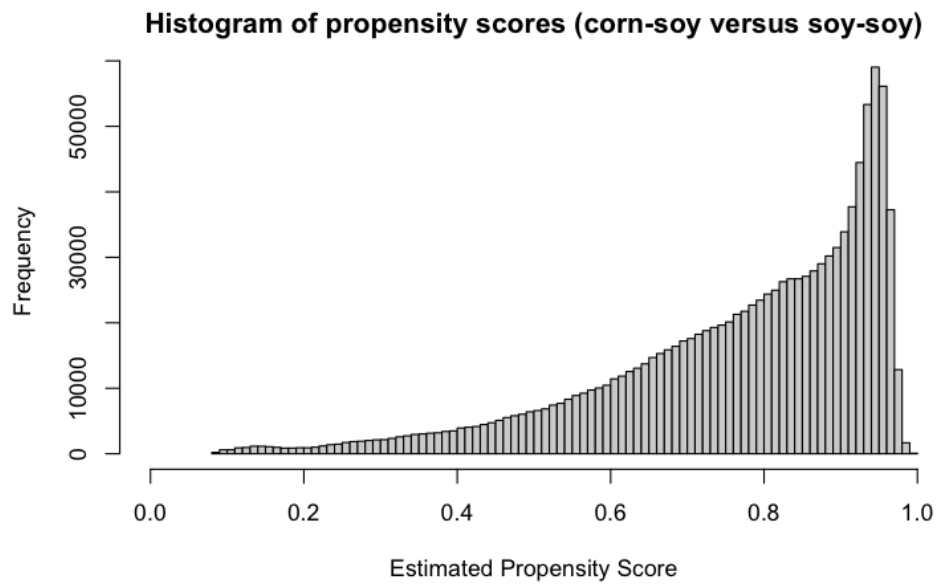

**Figure S1:** Top: Histogram of the propensity scores for the points used to train the causal forest that was used to calculate ECRB$_{Sat}$. The propensity score denotes the random forest-based estimated probability that a pixel growing corn grew soy in the previous year rather than corn. Bottom: Histogram of the propensity scores for the points used to train the causal forest that was used to calculate ESRB$_{Sat}$. The propensity score denotes the random forest-based estimated probability that a pixel growing soy grew corn in the previous year rather than soy.

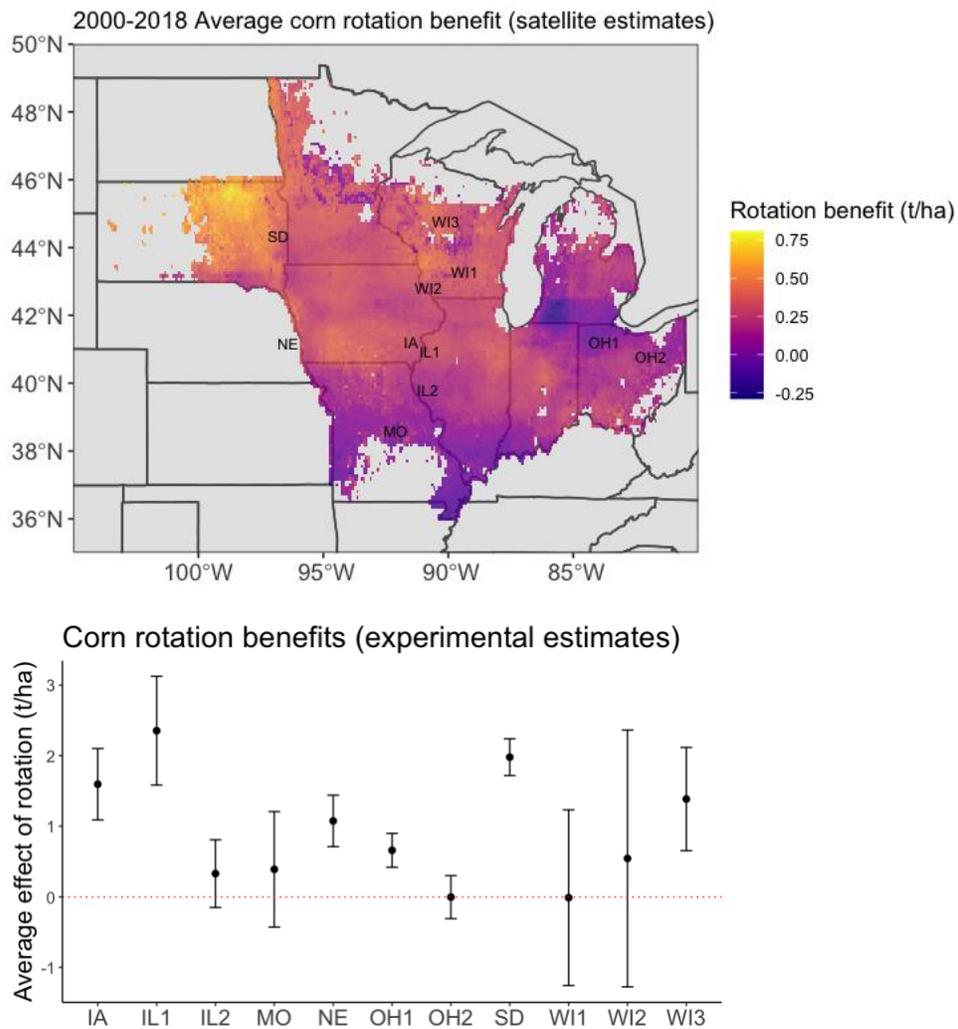

**Figure S2**: Top: ECRB$_{Sat}$ averaged across the years 2000-2018 and across 10 km$^2$ square bins. The text on the map indicates the locations of the experimental sites. Bottom: The average value across years of the experimental effect of rotation on corn yields, ECRB$_{Exp}$, at each of the 11 experimental sites. The error bars give 95 percent confidence intervals for the mean effect of rotation on yields at each site. The confidence intervals were calculated using site-specific standard errors of the difference between paired (rotation with non-rotation) subplots, and the standard errors were not pooled across sites.

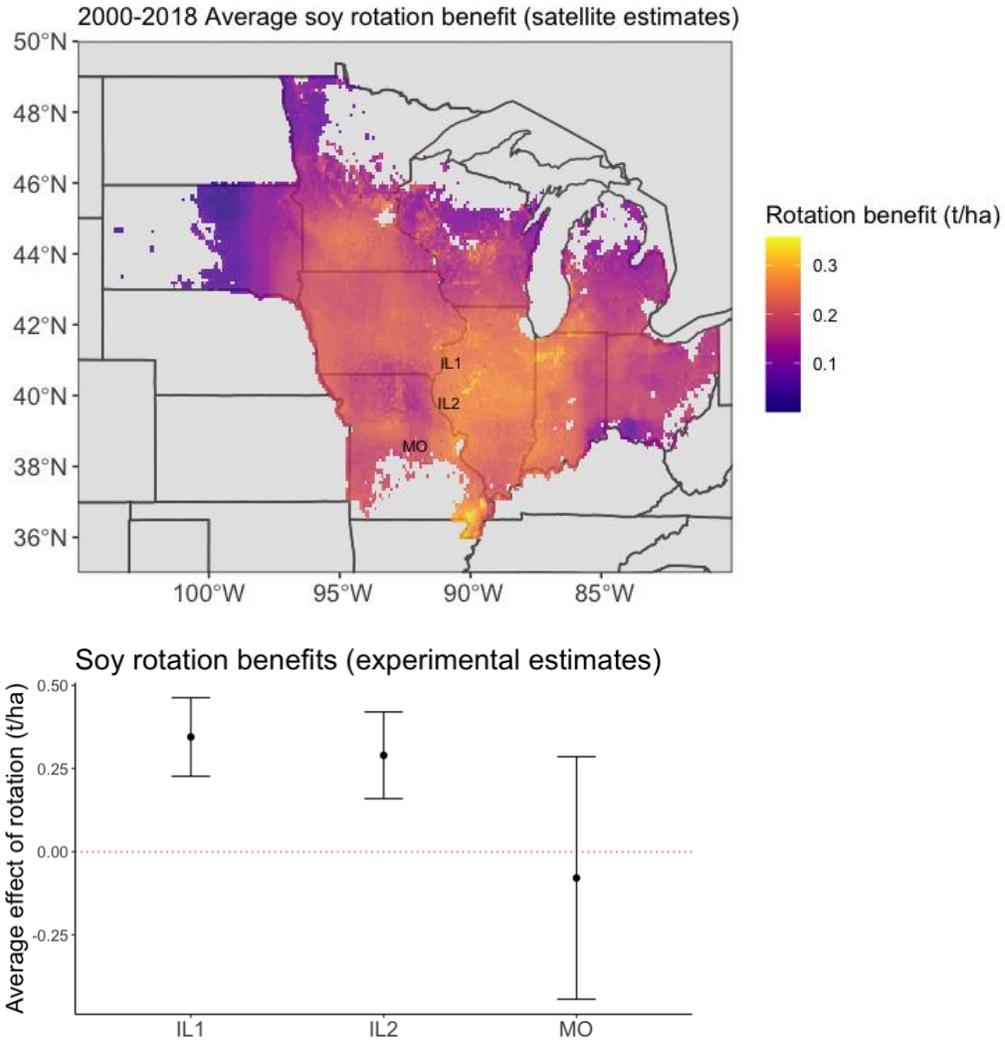

**Figure S3:** Top: ESRB$_{Sat}$ averaged across the years 2000-2018 and across 10 km$^2$ square bins. The text on the map indicates the locations of the experimental sites that had continuous soy rotations. Bottom: The average value across years of the experimental effect of rotation on soy yields, ESRB$_{Exp}$, at the 11 experimental sites. The error bars give 95 percent confidence intervals for the mean effect of rotation on soy yields at each site. The confidence intervals were calculated using site-specific standard errors of the difference between paired (rotation with non-rotation) subplots, and the standard errors were not pooled across sites.

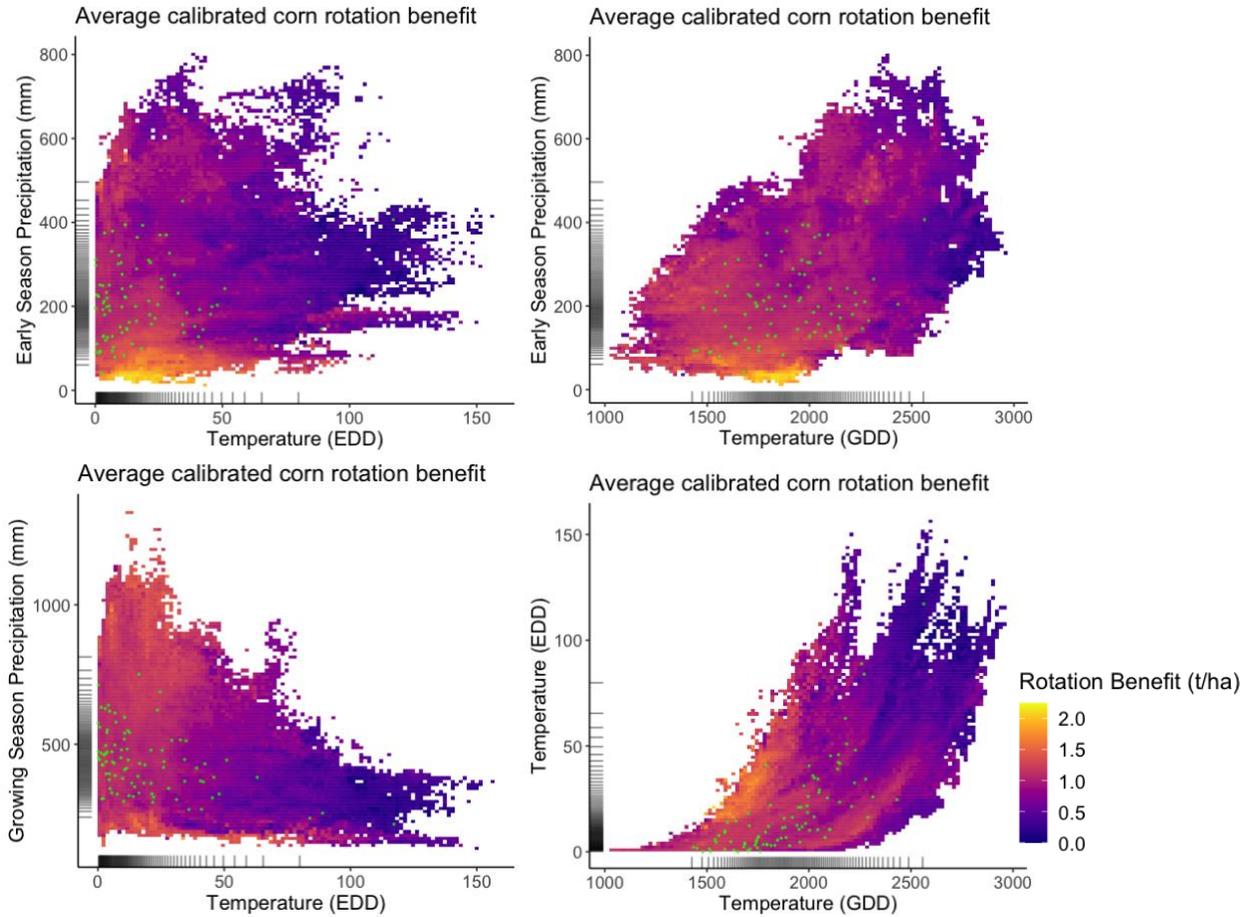

**Figure S4:** Heatmaps indicating how the calibrated corn rotation benefit, $ECRB_{Calib}$, varies with weather covariates. The green dots indicate the weather covariates of the experimental sites, with one dot per site per year. The 99 tick marks on each axes denote the $1^{st}$-$99^{th}$ percentiles of the corresponding weather covariate.

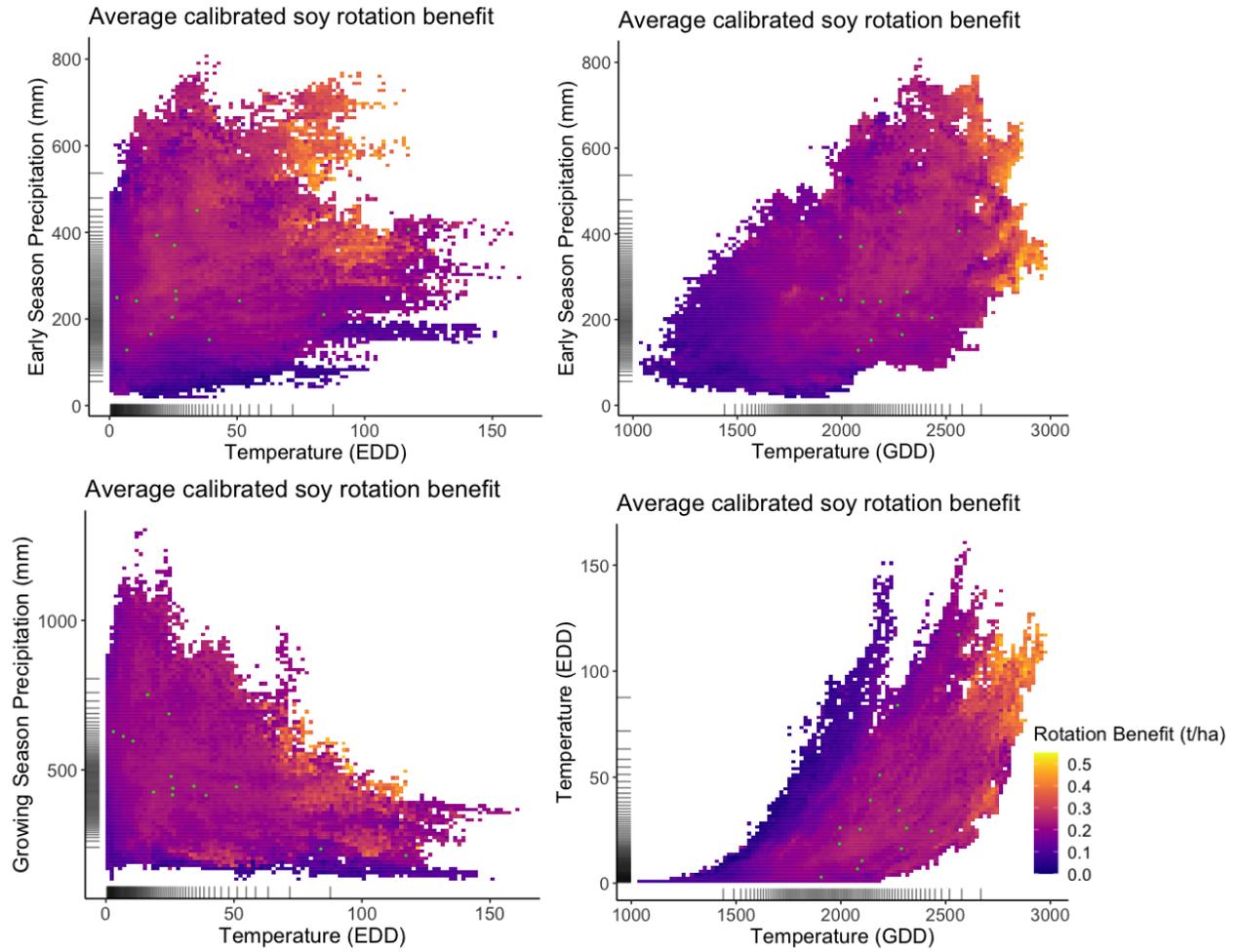

**Figure S5:** Heatmaps indicating how the estimated soy rotation benefit, ESRB$_{Calib}$, varies with weather covariates. The green dots indicate the weather covariates of the experimental sites that had continuous soy, with one dot per site per year. The 99 tick marks on each axes denote the 1$^{st}$-99$^{th}$ percentiles of the corresponding weather covariate.

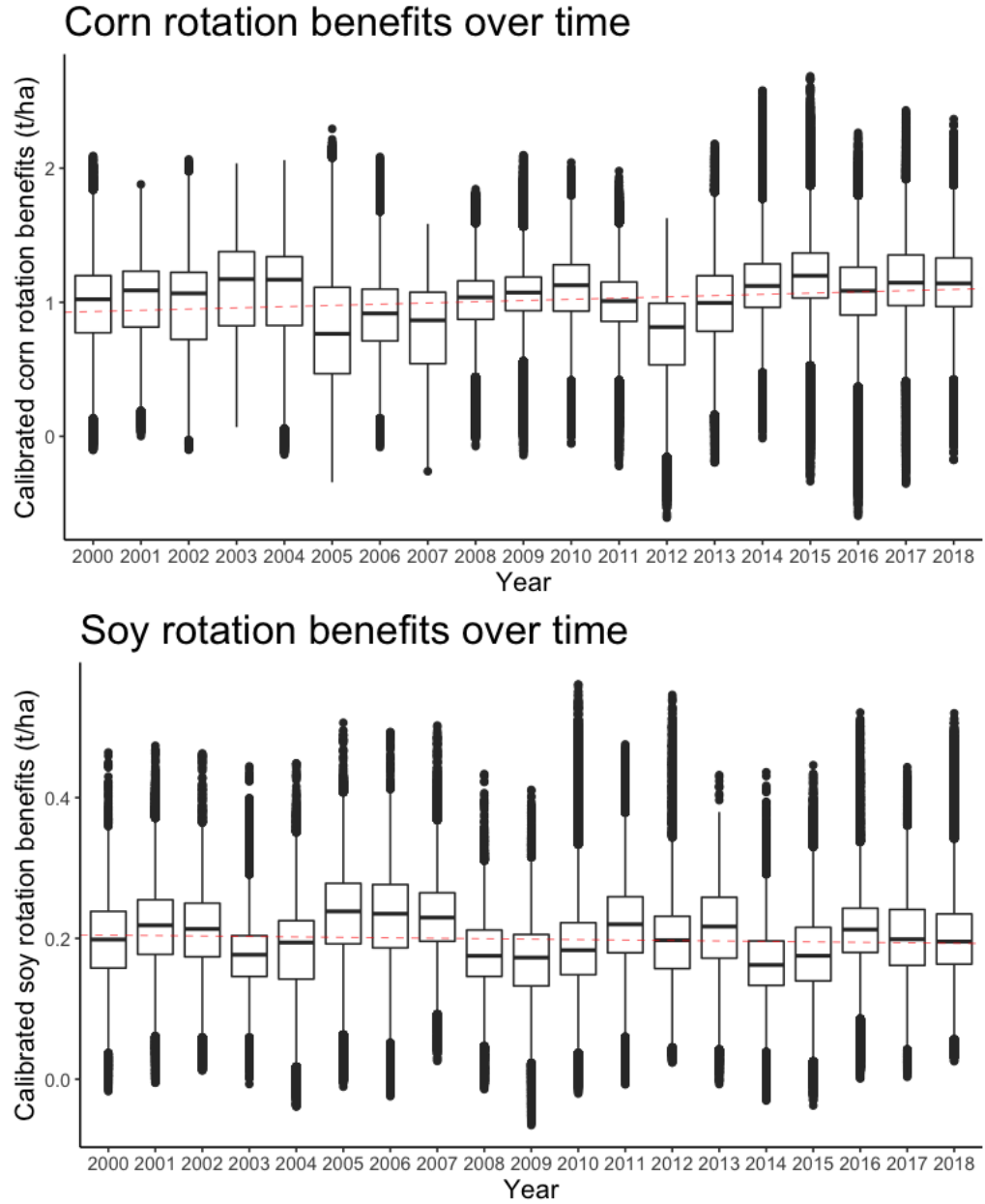

**Figure S6:** Boxplots of ECRB$_{Calib}$ values (top) and ESRB$_{Calib}$ values (bottom) by year. The red dotted lines are the best fit line using a standard linear model.

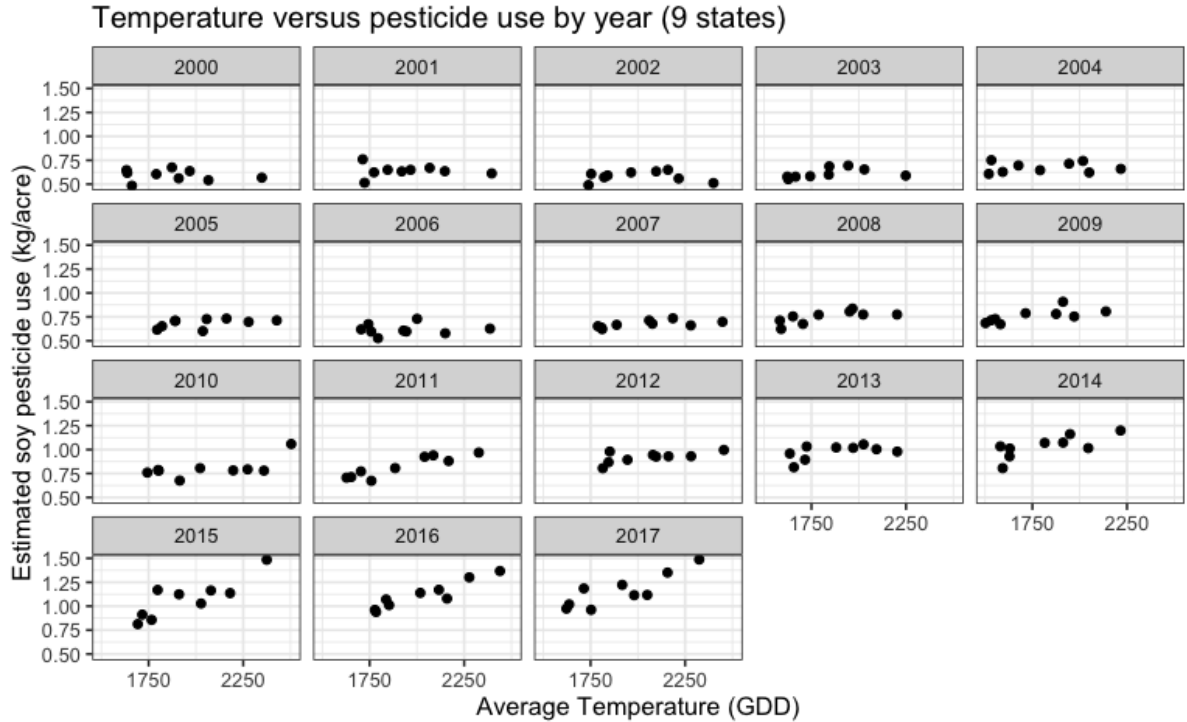

**Figure S7:** Average temperature in GDD versus estimated pesticide use for soy plotted by year. Each point corresponds to one of the 9 states in our study region (IA, IL, IN, MI, MN, MO, OH, SD, and WI). Prior to 2008, there is no apparent relationship between average temperatures in each state and soy pesticide use in that state. After 2007, it appears that higher temperatures are associated with increased pesticide use for soybean.

# Appendix A

Letting $T(t)$ be the temperature in Celsius as a function of time (in days), letting $t_1$ denote April 1st and letting $t_2$ denote September 30th, we used the following definition of extreme degree days

$$\text{EDD} = \int_{t_1}^{t_2} \max\{T(t) - 30, 0\}\, dt,$$

and the following definition of growing degree days,

$$\text{GDD} = \int_{t_1}^{t_2} \max\{T(t) - 8, 0\}\, dt - \int_{t_1}^{t_2} \max\{T(t) - 30, 0\}\, dt.$$

These metrics were chosen because corn growth is thought to benefit from higher temperatures in the 8° C to 30° C range, but is thought to suffer from temperature increases when the temperature exceeds 30° C. Therefore, high GDD is expected to benefit corn growth whereas high EDD is expected to damage corn growth.

To actually calculate the growing degree days (GDD) and extreme degree days (EDD) in our dataset, we used the daily minimum and maximum temperatures from GRIDMET between April 1st and September 30th. In particular, for each day and pixel, we considered the sine curve determined by the daily minimum and maximum temperature. The daily GDD and EDD contributions were estimated by applying a Riemann sum to the fitted sine curve with partition interval widths of one hour. The GDD and EDD were then estimated by summing the estimated GDD and EDD contributions across each day between April 1st and September 30th.

# **Appendix B**

Appendix B can be found at the following link:
https://drive.google.com/file/d/134Qj4EidoMFHqVg_L7WkMSUHvQJ5RUy6/view?usp=sharing

# Appendix C

Based on Figure S1 (bottom), we conclude that when estimating the causal effect of rotation on soy yields, the overlap assumption is not met. In particular, there are parts of covariate space where only CS rotations are observed and no SS rotations are observed as evidenced by the large concentration of points with propensity scores near 1. The $ESRB_{Sat}$ values corresponding to points with high propensity scores are prone to greater inaccuracies because they are estimated using few or no local observations (in covariate space) that had continuous soy. A common approach to dealing with this issue is to drop observations with propensity scores near 0 and 1. In this appendix, we drop $ESRB_{Sat}$ values for all observations where propensity score is less than 0.9, and ascertain that doing so does not give qualitatively different results than those in the main text. We do not drop points with propensity scores near 0, because those do not occur in our dataset.

In Figure S8, we check that dropping points with high propensity scores (at least 0.9) does not drastically affect the conclusions from Figure 3, which plots the geographic distribution of $ESRB_{Calib}$. Even if we had chosen to drop points with high propensity scores, we would still conclude a similar pattern of how the rotation benefit on soy yields varies geographically. While differences between the plots do exist (for example, in northwest Iowa), in both plots we see that the rotation benefit is highest in the southern and central parts of the Corn Belt.

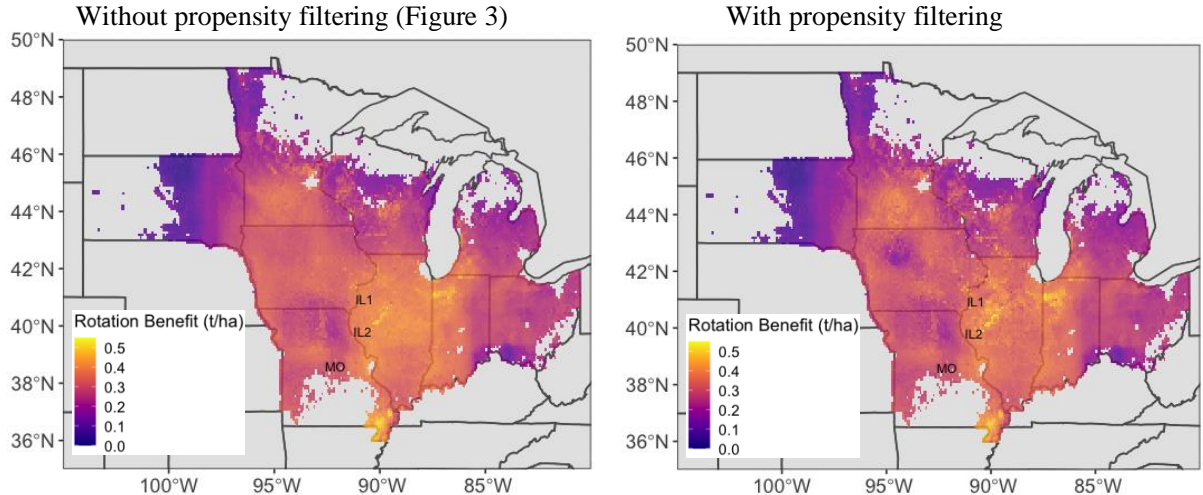

**Figure S8:** $ESRB_{Calib}$ averaged across the years 2000-2018 and across 10 km² square bins. The text on these maps indicate the locations of the experimental sites. The map on the left is the same as the one in Figure 3 of the main text. The map on the right is constructed in the same way except points where the causal forest estimated a propensity score of at least 0.9 were dropped from the analysis.

In Figure S9, we see that the conclusions that can be drawn from the heatmap of rotation benefits are not sensitive to our choice to keep the points with high propensity scores. In particular, we still observe that the rotation benefits increase with temperature when we remove points with high propensity scores from the analysis.

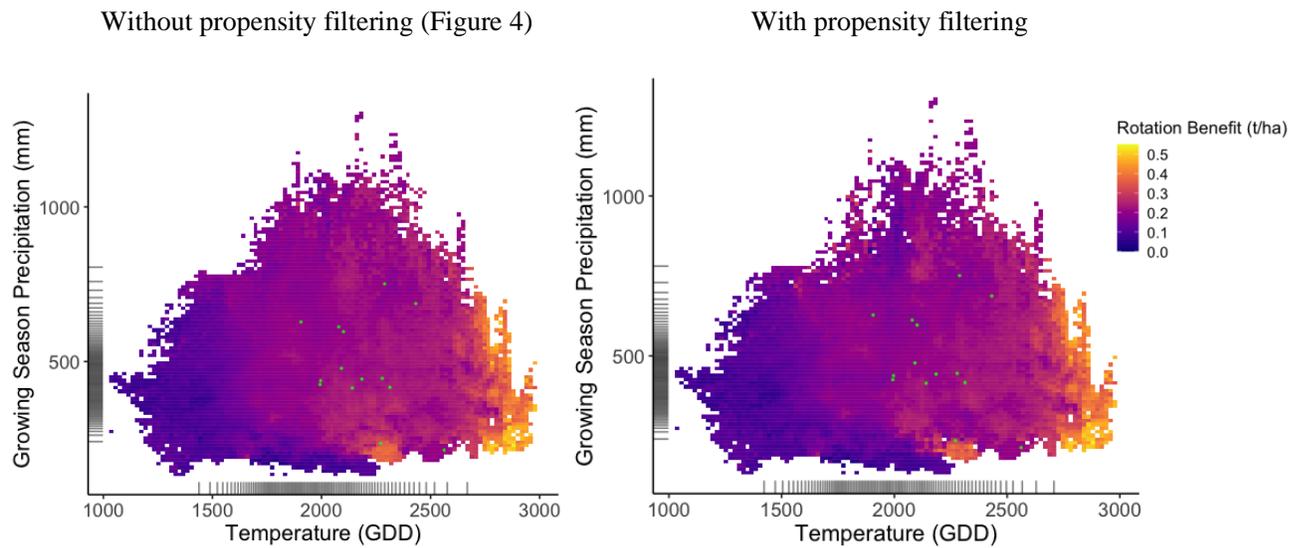

**Figure S9:** Heatmaps indicating how the calibrated soy rotation benefit, ESRB$_{Calib}$, varies with temperature and precipitation. The heatmap on the left is the same as the one in Figure 4 of the main text. The heatmap on the right is constructed in the same way except points where the causal forest estimated a propensity score of at least 0.9 were dropped from the analysis.

The numerical quantities reported in the main text are also not sensitive to our choice to not drop points with propensity scores greater than 0.9. In particular, we report in the main text average calibrated soy rotation benefit was 0.23 t/ha versus 0.15 t/ha across observations in the top and bottom quintile of GDD values, respectively. If we only consider points where the propensity score is at most 0.9, then the average calibrated soy rotation benefit become 0.23 t/ha and 0.14 t/ha across observations in the top and bottom quintile of GDD values, respectively. Additionally, we report in the main text that the overall average of the soy rotation benefit across the whole dataset is 0.20 t/ha, but if we only consider points where the propensity score is at most 0.9, then the average calibrated soy rotation benefit across the whole dataset is 0.19 t/ha.